\documentclass[10pt,journal]{IEEEtran}

\usepackage{times}
\usepackage{color}
\usepackage{soul}
\usepackage{url}
\usepackage[hidelinks]{hyperref}
\usepackage[utf8]{inputenc}
\usepackage[small]{caption}
\usepackage{graphicx}
\usepackage{amsmath}
\usepackage{booktabs}
\usepackage{algorithm}
\usepackage{algorithmic}
\newcommand{\natalia}[1]{{\color{black} #1}}
\title{Bias and Discrimination in AI: a cross-disciplinary perspective}

\author{\IEEEauthorblockN{Xavier Ferrer\IEEEauthorrefmark{1},
Tom van Nuenen\IEEEauthorrefmark{1},
Jose M. Such\IEEEauthorrefmark{1}, 
Mark Cot\'e\IEEEauthorrefmark{1}, and
Natalia Criado\IEEEauthorrefmark{1}},

\IEEEauthorblockA{\IEEEauthorrefmark{1}King's College London, United Kingdom}

\thanks{
Corresponding author: X. Ferrer (email: xavier.ferrer\_aran@kcl.ac.uk).}}

\begin{document}

\maketitle

\begin{abstract}
With the widespread and pervasive use of Artificial Intelligence (AI) for automated decision-making systems, AI bias is becoming more apparent and problematic. One of its negative consequences is discrimination: the unfair, or unequal treatment of individuals based on certain characteristics. However, the relationship between  bias and discrimination is not always clear. 
In this paper, we survey relevant literature about bias and discrimination in AI from an interdisciplinary perspective that embeds technical, legal, social and ethical dimensions. We show that finding solutions to bias and discrimination in AI requires robust cross-disciplinary collaborations.
\end{abstract}

%
%

\section{Introduction}
Operating at a large scale and impacting large groups of people, automated systems can make consequential and sometimes contestable decisions. Automated decisions can impact a range of phenomena, from credit scores to insurance payouts to health evaluations. 
These forms of automation can become problematic when they place certain groups or people at a systematic disadvantage. These are cases of discrimination – which is legally defined as the unfair or unequal treatment of an individual (or group) based on certain characteristics such as income, education, gender or ethnicity.
When the unfair treatment is caused by automated decisions, usually taken by intelligent agents or other AI-based systems, we talk about digital discrimination. Digital discrimination has been found in a diverse range of fields, such as in risk assessment systems for policing and credit scores \cite{o2017weapons}.

Digital discrimination is becoming a serious problem, as more and more decisions are delegated to systems increasingly based on AI techniques such as Machine Learning. While a significant amount of research has been undertaken from different disciplinary angles to understand this challenge – from computer science to law to sociology – none of these fields have been able to resolve the problem on their own terms. For instance, computational methods to verify and certify bias-free datasets and algorithms do not account for socio-cultural or ethical complexities, and do not distinguish between bias and discrimination. Both of these terms have a technical inflection, but are predicated on legal and ethical principles.

In this paper, we propose a synergistic approach that allows us to explore bias and discrimination in AI by supplementing technical literature with social, legal and ethical perspectives. Through a critical survey of a synthesis of related literature, we compare and evaluate the sometimes contradictory priorities within these fields, and discuss how disciplines might collaborate to resolve the problem. We also highlight a number of interdisciplinary challenges to attest and address discrimination in AI.

\section{Bias and Discrimination}\label{sec:back}

Technical literature in the area of discrimination typically refers to the related issue of bias. Yet, despite playing an important role in discriminatory processes, bias does not necessarily lead to discrimination.  Bias means a deviation from the standard, sometimes necessary to identify the existence of some statistical patterns in the data or language used \cite{Danks2017, ferrer2020discovering}. Classifying and finding differences between instances would be impossible without bias. 

In this paper, we follow the most common definition of bias used in the literature and focus on the \emph{problematic} instances of bias that may lead to discrimination by AI-based automated-decision making systems.
Three main, well-known causes for bias have been distinguished \cite{Danks2017}:

\paragraph{Bias in modelling}
Bias may be deliberately introduced, e.g., through smoothing or regularisation parameters to mitigate or compensate for bias in the data, which is called \emph{algorithmic processing bias}, or introduced while modelling in cases with the usage of objective categories to make subjective judgements, which is called \emph{algorithmic focus bias}. 

\paragraph{Bias in training} 
Algorithms learn to make decisions or predictions based on datasets that often contain past decisions. If a dataset used for training purposes reflects existing prejudices, algorithms will very likely learn to make the same biased decisions. Moreover, if the data does not correctly represent the characteristics of different populations, representing an \emph{unequal ground truth}, it may result in biased algorithmic decisions.

\paragraph{Bias in usage} 
Algorithms can result in bias when they are used in a situation for which they were not intended. An algorithm utilised to predict a particular outcome in a given population can lead to inaccurate results when applied to a different population – a form of \emph{transfer context bias}. Further, the potential misinterpretation of an algorithm's outputs can lead to biased actions through what is called \emph{interpretation bias}.

A significant amount of literature focuses on forms of bias that may or may not lead to discriminatory outcomes, i.e., the \textit{relationship} between bias and discrimination is not always clear or understood. Most literature assumes that systems free from biases do not discriminate, \natalia{hence,} reducing or eliminating biases reduces or eliminates the potential for discrimination. However, whether an algorithm can be considered discriminatory or not depends on the context in which it is being deployed and the task it is intended to perform. For instance, consider a possible case of algorithmic bias in usage, in which an algorithm is biased towards hiring young people. At first glance, it can be considered that the algorithm is discriminating against older people. However, this (biased) algorithm should only be considered to discriminate if the context in which it is intended to be deployed does not justify hiring more young people than older people. Therefore, statistically reductionist approaches, such as estimating the ratio between younger and older people hired, are insufficient to attest whether the algorithm is discriminating without considering this socially and politically fraught context; 
it remains ethically unclear where we need to draw the line between biased and discriminating outcomes. Therefore, AI and technical researchers often: i) use discrimination and bias as equivalent; or ii) focus on measuring biases without actually attending to the problem of whether or not there is discrimination. 
Our aim, in the below, is to disentangle some of these issues.

%
%
%
%

\section{Measuring Biases}\label{sec:tech}

To assess whether an algorithm is free from biases, there is a need to analyse the entirety of the algorithmic process. 
This entails first confirming that the algorithm's underlying assumptions and its modelling are not biased; second, that its training and test data does not include biases and prejudices; and finally, that it is adequate to make decisions for that specific context and task. More often than not, however, we do not have access to this information. 
A number of issues prevent such an analysis. The data used to train a model, for instance, is typically protected since it contains personal information, rendering the task of attesting training bias impossible. Access to the algorithm's source code might also be restricted to the general public, removing the possibility of identifying modelling biases. This is common as algorithms are valuable private assets of companies. Third, the specifics of where and how the algorithm will be deployed might be unknown to an auditor. Depending on what is available, different types of bias attesting might be possible, both in terms of the process and in terms of the metrics used to measure it.

\subsection{Procedural vs Relational Approaches}\label{sec:fairnessbiasdiscr}

We can distinguish between two general approaches to measure bias: i) procedural approaches, which focus on identifying biases in the decision making process of an algorithm \cite{Mueller2019}, and ii) relational approaches, which focus on identifying (and preventing) biased decisions in the dataset or algorithmic output. While ensuring unbiased outcomes is useful to attest whether a specific algorithm has a discriminatory impact on a population, focusing on the algorithmic process itself can help yield insights about the reason why it happened in the first place.

Procedural approaches focus on identifying biases in the algorithmic ``logic''. Such ante-hoc interventions are hard to implement for two main reasons: (i) AI algorithms are often sophisticated and complex since, in addition to being trained on huge data sets, they usually make use of unsupervised learning structures that might prove difficult to trace and understand (e.g. neural networks), and (ii) the source code of the algorithm is rarely available.
Procedural approaches will become more beneficial with further progress in explainable AI \cite{Mueller2019}. 

Being able to understand the process behind an algorithmic discriminatory decision can help us understand possible problems in the algorithm's code and behaviour, and thus act accordingly towards the creation of non-discriminatory algorithms. As such, current literature on non-discriminatory AI promotes the introduction of explanations into the model itself, e.g., through inherently interpretable models such as decision trees, association rules, or causal reasoning which provide coarse approximations of how a system behaves by explaining the weights and relationships between variables in (a segment of) a model \cite{guidotti2018survey,Ruggieri2010,kilbertus2017avoiding}. Notice, however, that attesting that an algorithmic process is free from biases does not ensure a non-discriminatory algorithmic
output, since discrimination can arise as a consequence of
biases in training or in usage \cite{calders2013unbiased}.

While procedural approaches attend to the algorithmic process, relational approaches measure biases in the dataset and the algorithmic output. Such approaches are popular in the literature, as they do not require insights into the algorithmic process. Besides evaluating biases in the data itself, where it is available (e.g. by looking at statistical parity), implementations can compare the algorithmic outcomes obtained by two different sub-populations in the dataset \cite{criado2020normative}, or make use of counterfactual or contrastive explanation, asking questions such as ``Why X instead of Y?''. Bias, here, is only located at testing time. One example is the post-hoc approach of Local Interpretable Model-Agnostic Explanations (LIME), which makes use of adversarial learning to generate counterfactual explanations \cite{Mueller2019}.
Other approaches evaluate the correlation between algorithmic inputs 
and biased outputs, in order to identify those features that may lead to biased actions that affect protected sub-populations \cite{Grgic-Hlaca2018}. Since implementations often ignore the context in which the algorithm will be deployed, the decision whether a biased output results in a case of discrimination is often left to the user to assess \cite{Ruggieri2010}. 


\subsection{Bias Metrics}\label{sec:metrics}
The metrics for measuring bias can be organised in three different categories: statistical measures, similarity-based measures, and causal reasoning. While reviews such as \cite{verma2018fairness} offer an extensive description of some of these metrics, we will discuss the intuition behind the most common types of metrics used in the literature below. 

Statistical measures to attest biases represent the most intuitive notion of bias, and focus on exploring the relationships or associations between the algorithm's predicted outcome for the different (input) demographic distributions of subjects, and the actual outcome that is achieved. These measures include, first, \emph{group fairness} (also named  \emph{statistical parity}), which requires that an equal quantity of each group of distinct individuals should receive each possible algorithmic outcome. For instance, if four out of five applicants of the advantaged group were given a mortgage, the same ratio of applicants from the protected group should obtain the mortgage as well. Second, \emph{predictive parity} is satisfied if both protected and unprotected groups have equal positive predictive value – that is, the probability of an individual to be correctly classified as belonging to the positive class. Finally, the principle of \emph{well-calibration} states that the probability estimates provided by the decision-making algorithm should be properly adjusted with the real values. 
Despite the popularity of statistical metrics, it has been shown that statistical definitions are insufficient to estimate the absence of biases in algorithmic outcomes, as they often assume the availability of verified outcomes necessary to estimate them, and often ignore other attributes of the classified subject than the sensitive ones \cite{dwork2012fairness}.

Similarity measures, on the other hand, focus on defining a similarity value between individuals. \emph{Causal discrimination} is an example of such measures, stating that a classifier is not biased if it produces the same classification for any two subjects with the same non-protected attributes. A more complex bias metric based on a similarity measure between individuals is \emph{fairness through awareness} \cite{dwork2012fairness}, which states that, for fairness to hold, the distance between the distributions of outputs for individuals should \textit{at most} be the distance between the two individuals as estimated by means of a similarity metric. The complexity in using this metric consists in accurately defining a similarity measure that correctly represents the complexity of the situation in question, which is often an impossible task to generalise. Moreover, the similarity measure between individuals can suffer from the implicit biases of the expert, resulting in a biased similarity estimator.

Finally, definitions based on causal reasoning assume bias can be attested by means of a directed causal graph. In the graph, attributes are presented as nodes joined by edges which, by means of equations, represent the relations between attributes \cite{kilbertus2017avoiding}. By exploring the graph, the effects that the different protected attributes have on the algorithm's output can be assessed and analysed. Causal fairness approaches are limited by the assumption that a valid causal graph able to describe the problem can be constructed, which is not always feasible due to the sometimes unknown and complex relations between attributes and the impact they have on the output.

\section{Attesting and Addressing Discrimination}\label{sec:iob}

The first step explored in the related literature to identify discriminatory outputs is determining the groups whose algorithmic outputs are going to be compared. Technical approaches to select the sub-populations of interest vary, either: i) they consider sub-populations as already defined \cite{Ruggieri2010,Feldman2014}; or ii) they are selected by means of a heuristic that aggregates individuals that share one or more protected or proxy attributes, as in \emph{FairTest}'s framework\footnote{\url{https://github.com/columbia/fairtest}} for detecting biases in datasets. \emph{Protected attributes} are encoded in legislation (cf.~Sect.~\ref{sec:legal}) and usually include attributes such as sex, gender, and ethnicity, while \emph{proxy attributes} are attributes strongly correlated with protected attributes, e.g. weightlifting ability (strongly correlated with gender). However, the process of selecting individuals or groups based on these attributes is non-trivial since groups often result from the intersection of multiple protected and proxy attributes (cf.~Sect.~\ref{sec:social}).

Once the protected and the potentially advantaged groups have been selected, implementations apply different bias metrics (cf.~Sect.~\ref{sec:metrics}) to compare and identify relevant differences in the algorithm's outcomes for the different groups. If these differences are a consequence of protected attributes, it is \emph{likely} that the algorithm's decision can be considered discriminatory. 

To alleviate the contextual problem of whether an algorithmic outcome may form a case of discrimination, approaches often incorporate \emph{explanatory attributes}: user attributes on which is deemed acceptable to differentiate, even if this leads to apparent discrimination on protected attributes \cite{Feldman2014}. Some relevant approaches are the open-source IBM AI Fairness 360 toolkit\footnote{https://github.com/IBM/AIF360}, which contains techniques developed by IBM and the research community to help detect and mitigate bias in machine learning models throughout the AI application lifecycle, and Google's What-if-tool\footnote{\url{https://pair-code.github.io/what-if-tool/}}, which offers an interactive visual interface that allows researchers to investigate model performances for a range of features in the dataset and optimization strategies.

Despite these efforts in parameterising context uncertainty in technical implementations, the interpretive dimension that separates bias and discrimination remains a challenge. As a response, some approaches base their implementations on various anti-discrimination laws that focus on the relationships between protected attributes and decision outcomes. For instance, the US \emph{fourth-fifth court rule} and the \emph{Castaneda rule} are used as a general, and often arguably adequate, \emph{prima facie} evidence of discrimination – see Section \ref{sec:legal} for more details on these rules. 

Approaches that intervene on problematic biases focus on (i) removing protected attributes from the data, as an attempt to impede the algorithm from using these protected attributes to make discriminatory decisions (\emph{fairness through blindness} \cite{dwork2012fairness,calders2013unbiased}), or on (ii) debiasing algorithms' outputs \cite{bolukbasi2016man}. An issue here is that removing protected attributes from the input data often results in a significant loss of accuracy in the algorithm \cite{dwork2012fairness}. Moreover,  excluded attributes can often be correlated with \emph{proxy attributes} that remain in the dataset, meaning bias may still be present (i.e. certain residential areas have specific demographics that play the role of proxy variables for ethnicity. These approaches can also be criticised because they alter the model of the world that an AI makes use of, instead of altering how that AI perceives and acts on bias \cite{dwork2012fairness}. 

On a broader level, debiasing an algorithm's output requires a specific definition of its context and, as such, is difficult to achieve from a technical perspective only. A myriad of lingering questions remains to be answered: how \emph{much} bias does an algorithm need to encode in order to consider its outputs discriminating? 
How can we reflect on the peculiarity of the data on which these algorithms are operating – data which often reflects the inequities of its time? In short, a clearer definition of the relation between algorithmic biases and discrimination is needed. We argue that such a definition can only be provided by a cross-disciplinary approach that takes legal, social and ethical considerations into account. In response, in the next sections we will engage critically with related work from legal, social and ethical perspectives.

%
%
%
%
\section{Legal Perspective }\label{sec:legal}

Legislation designed to prevent discrimination against particular groups of people that share one or more protected attributes – namely \emph{protected groups} – receives the name of anti-discrimination law. Anti-discrimination laws vary across countries. For instance, European anti-discrimination legislation is organised in directives, such as Directive 2000/43/EC against discrimination on grounds of race and ethnic origin, or Chapter 3 of the EU Charter of fundamental rights. Anti-discrimination laws in the US are described in the \emph{Title VII of the Civil Rights Act of 1964} and in other federal and state statutes, supplemented by court decisions. For instance, the Title VII prohibits discrimination in employment on the basis of race, sex, national origin and religion; and the \emph{The Equal Pay Act} prohibits \natalia{wage disparity} based on sex by employers and unions.

The main issues in trials related to discrimination consist of determining \cite{romei2014multidisciplinary}: (1) the relevant population affected by the discrimination case, and to which groups it should be compared, (2) the discrimination measure that formalises group under-representation, e.g., \emph{disparate treatment} or \emph{disparate impact} \cite{Feldman2014,Barocas2016}, and (3) the threshold that constitutes prima facie evidence of discrimination. Note that the three issues coincide with the problems explored in the technical approaches presented earlier. With respect to the last point, no strict threshold has been laid down by the European Union. In the US, the \emph{fourth-fifth rule} from the Equal Employment Opportunity Commission (1978), which states that a job selection rate for the protected group of less than 4/5 of the selection rate for the unprotected group, is sometimes used a prima facie evidence of an adverse impact. The \emph{Castaneda rule}, which states that the number of people of the protected group selected from a relevant population cannot be smaller than 3 standard deviations the number expected in a random selection, is also used \cite{Barocas2016}. While such laws can relieve discriminatory issues, more complex scenarios can arise.  For instance, Hildebrandt and Koops mention the legally grey area of price discrimination, where consumers in different geographical areas can be offered different prices based on differences in average income \cite{Hildebrandt2010}.

More recent regulations, such as the General Data Protection Regulation (GDPR), have been offered as a framework to alleviate some of the enforcement problems of anti-discrimination law, and include clauses on automated decision-making related to \emph{procedural regularity} and accountability, introducing a right of explanation for all individuals to obtain meaningful explanations of the logic involved when automated decision making takes place. However, these solutions often assume white box scenarios, which, as we have seen, may be difficult to achieve technically, and even when they are achieved, they may not necessarily provide the answers sought to assess whether discrimination is present or not. Generally speaking, current laws are badly equipped to address algorithmic discrimination \cite{Barocas2016}. Leese \cite{Leese2014}, for instance, notes that anti-discrimination frameworks typically follow the establishment of a causal chain between indicators on the theoretical level (e.g. sex or race) and their representation in the population under scrutiny. Data-driven analytics, however, create aggregates of individual profiles, and as such are prone to the production of arbitrary categories instead of real communities. As such, even \textit{if} data subjects are granted procedural and relational explanations, the question remains at which point potential biases can reasonably be considered forms of discrimination. 

%
%
%
%
\section{Social Perspective }\label{sec:social}

Digital discrimination is not only a technical phenomenon regulated by law, but one that also needs to be considered from a socio-cultural perspective in order to be rigorously understood. Defining what constitutes discrimination is a matter of understanding the particular social and historical conditions and ideas that inform it, and needs to be reevaluated according to its implementation context.
Bias in usage, as defined above, forms a challenge to any kind of generalist AI solution.

One complication highlighted by a social perspective is the potential of digital discrimination to reinforce existing social inequalities. This point becomes increasingly pressing when multiple identities and experiences of exclusion and subordination start interacting – a phenomenon called intersectionality \cite{Walby2012}. One example is formed by the multiple ways that race and gender interact with class in the labour market, effectively generating new identity categories. From a legislation perspective, anti-discrimination laws can be applied when discrimination is experienced by a population that shares one or more protected attributes. However, this problem can exponentially grow in complexity when also considering proxy variables and the intersection of different features \cite{Grgic-Hlaca2018}.

On a cultural and ideological level, the call for ever-expanding transparency of AI systems needs to be seen as an \emph{ideal} as much as a form of 'truth production'
\cite{Ananny2018}. Further, no standard evaluation methodology exists among AI researchers to ethically assess their bias classifications, as the explanation of classification serves different functions in different contexts, and is arguably assessed differently by different people (for instance, the way a dataset is defined and curated, for instance, depends on the assumptions and values of the creator) \cite{vantransparency}. 
Conducting a set of experimental studies to elicit people’s responses to a range of algorithmic decision scenarios and explanations of these decisions, \cite{Binns2018} find a strong split in their respondents: some find the general idea of algorithmic discrimination immoral, others resist imputing morality to a computer system altogether \emph{'the computer is just doing its job'}~\cite{Binns2018}. While algorithmic decision-making implicates dimensions of justice, its claim to objectivity may also preclude the public awareness of these dimensions.

Given the differing stances on discrimination in society, providing explanations to the public targeted by algorithmic decision-making systems is key, as it allows individuals to make up their own minds about their evaluations of these systems. Hildebrand and Koops in \cite{Hildebrandt2010}, for instance, call for \emph{smart transparency} by designing the socio-technical infrastructures responsible for decision-making in a way that allows individuals to anticipate and respond to how they are profiled. 
In this context of public evaluation, it also becomes important to question which moral standards can or should be encoded in AI, and which considerations of discrimination can be expected to be most readily shared by a widely differing range of citizens \cite{Curry2019}. While such frameworks can always be criticised as reductionist approaches to the complexity of social values, keeping into account what kinds of values are important in society can go some way in helping to establish \emph{how} discrimination can be defined. 

%
%

\section{Ethical Perspective}\label{sec:ethics}

Finally, we need to bring in ethical perspective; as Tasioulas argues, discrimination does not need to be unlawful in order to be unfair \cite{tasioulas2018first}. Yet, moral standards are historically dynamic, and continuously evolving due to technological developments. This explains why law and encoded social morality often lag behind technical developments. In light of discriminatory risks (and benefits) that AI might pose, moral standards need to be reassessed in order to enable new definitions of discriminatory impact. It is telling that one of the famous attempts to address this question in robotics derives from fiction: Isaac Asimov’s Three Laws of Robotics. More recently, the AI community has attempted to codify ethical principles for AI, such as the Asilomar AI Principles\footnote{https://futureoflife.org/ai-principles/}. However, these principles are criticised as being vague, mainly due to their level of abstraction, making them not necessarily helpful \cite{tasioulas2018first}. 

More grounded and detailed frameworks for AI ethics have recently been proposed, such as the standards being defined by the IEEE Global Initiative on Ethics of Autonomous and Intelligent Systems\footnote{\url{https://ethicsinaction.ieee.org/}}, which aim to provide an incubation space for new solutions relevant to the ethical implementation of intelligent technologies. Another noteworthy contribution is presented in \cite{tasioulas2018first}, stating that the ethical questions related to the usage of AI can be organised into three interconnected levels. The first level involves laws to govern AI-related activities, including public standards backed up by public institutions and enforcement mechanisms, which claim to be morally binding on all citizens in virtue of their formal enactment. Some efforts discussed in Section \ref{sec:legal} can be seen as examples of this. However, this evades the problem that not all of the socially entrenched standards that govern our lives are legal standards. We rely not only on the law to discourage people from wrongful behaviour, but also on moral standards that are instilled in us from childhood and reinforced by society. 

The second level is the social morality around AI. The definition of such a morality is problematic as it involves a potential infinity of reference points, as well as the cultivation of emotional responses such as guilt, indignation and empathy – both of which are effects of human consciousness and cognition \cite{tasioulas2018first}. The third and final level includes individuals and their engagement with AI. Individuals and associations will still need to exercise their own moral judgement by, for instance, devising their own codes of practice. However, how these levels can be operationalised (or to what extent) from a technical AI point of view is not yet clear.

%
%
%
%
\section{Open Challenges}

Addressing and attesting digital discrimination and remedying its corresponding deficiencies will remain a problem for technical, legal, social, and ethical reasons. 
Technically, there are a number of practical limits to what can be accomplished, particularly regarding the ability to automatically determine the relationship between biases and discrimination. Current legislation is poorly equipped to address the classificatory complexities arising from algorithmic discrimination. Social inequalities and differing attitudes towards computation further obfuscate the distinction between bias and discrimination. From an ethical perspective, existing moral standards need to be reassessed in light of the risks and benefits AI might pose. 

In sum, the design and evaluation of AI systems is rooted in different perspectives, concerns and goals. To posit the existence of a predefined path through these perspectives would be misleading. What is needed, instead, is a sensitivity to the distinctions concerning what is desirable AI implementation, and to a dialogical orientation towards design processes.
Finding solutions to discrimination in AI requires robust cross-disciplinary collaborations. We conclude here by summarising what we believe to be  some of the most important cross-disciplinary challenges to advance research and solutions for attesting and avoiding discrimination in AI.

\subsection{How Much Bias Is Too Much?}
Whether a biased decision can be considered discriminatory or not depends on many factors, such as the context in which AI is going to be deployed, the groups compared in the decision, and other factors like a trade-off between individualist-meritocratic and outcome-egalitarian values. To simplify these problems, technical implementations tend to borrow definitions from the legal literature, such as the thresholds that constitute prima facie evidence of discrimination, and use it as a general rule to attest algorithmic discrimination. Yet this cannot be addressed by simply encoding the legal, social and ethical context, which in and of itself is nontrivial. Bias and discrimination have a different ontological status: while the former may seem easy to define in terms of programmatic solutions, the latter involves a host of social and ethical issues that are challenging to resolve from a positivist framework.

\subsection{Critical AI Literacy}
Another challenge is the need for an improvement in critical AI literacy. We have noted the need to take into account the end user of AI decision making systems, and the extent to which their literacy of these systems can be targeted and improved. In part, this entails end user knowledge of particularities such as the attributes being used in a dataset, as well as the ability to compare explanation decisions and moral rules underlying those choices. This is, however, not solely a technical exercise, as decision making systems render end users into algorithmically constructed data subjects. This challenge could be addressed through a socio-technical approach which can consider both the technical dimensions and the complex social contexts in which these systems are deployed. Building public confidence and greater democratic participation in AI systems requires ongoing development of not just explainable AI but of better Human-AI interaction methods and socio-technical platforms, tools and public engagement to increase critical public understanding and agency.

\subsection{Discrimination-aware AI}
Third, AI should not just be seen as a potential problem causing discrimination, but also as a great opportunity to mitigate existing issues. The fact that AI can pick up on discrimination suggests it can be made \emph{aware} of it. For instance, AI could help spot digital forms of discrimination, and assist in acting upon it. For this aim to become a reality we would need, as explored in this work, a better understanding of social, ethical, and legal principles, as well as dialogically constructed solutions in which this knowledge is incorporated into AI systems.  Two ways to achieve this goal are: i) using data-driven approaches like machine learning to actually look at previous cases of discrimination and try to spot them in the future; and ii) using model-based and knowledge-based AI that operationalises the socio-ethical and legal principles mentioned above (e.g., normative approaches that include non-discrimination norms as part of the knowledge of an AI system to influence its decision making). This would, for instance, facilitate an AI system realising that the knowledge it gathered or learned is resulting in discriminatory decisions when deployed in specific contexts. Hence, the AI system could alert an expert human about this, and/or proactively address the issue spotted. 

\section*{ACKNOWLEDGMENT}
This work was supported by EPSRC under grant EP/R033188/1. It is part of the Discovering and Attesting Digital Discrimination (DADD) project – see https://dadd-project.org.
\bibliographystyle{IEEEtran}
\bibliography{main}

\end{document}